# The observation resolution - a neglected aspect with critical influence on movement-based foraging indices

Michael Kalyuzhny[1,a], Tom Haran[1,2] and Dror Hawlena[1,2]

[1] Department of Ecology, Evolution & Behavior, Alexander Silberman Institute of life Sciences, the Hebrew University of Jerusalem, Edmond J. Safra Campus, Jerusalem **91904**, Israel.

[2] Herpetological Collection, National Natural History Collections, the Hebrew University of Jerusalem, Jerusalem, Israel.

[a] Corresponding author: michael.kalyuzhny@mail.huji.ac.il

## Abstract

1. Movement-based indices such as moves per minute (MPM) and proportion time moving (PTM) are common methodologies to quantify foraging behaviour. Hundreds of studies have reported these indices without specifying the temporal resolution of their original data, despite the likelihood that the minimal stop and move durations can affect MPM and PTM estimates.
2. Our goal was to empirically determine the sensitivity of these foraging indices to changes in the temporal resolution of the observation.
3. We used a high speed camera to record movement sequences of 20 *Acanthodactylus boskianus* lizards. We gradually decreased the data resolution by ignoring short stops and either ignoring, elongating or leaving short moves unchanged. We then used the manipulated data to calculate the foraging indices at different temporal resolutions.
4. We found that movement-based indices are very sensitive to the observation resolution, so that realistic variation in the minimal duration of stops and moves could lead to 68% and 48% difference in MPM and PTM estimates, respectively. When using the highest resolution, our estimate of MPM was an order of magnitude higher than all prior reported values for lizards. Also, the distribution of stop durations was well described by a single heavy tailed distribution above 0.35 seconds. This suggests that for *A. boskianus* there is no reason to ignore short stops above this threshold.
5. Our results raise major concerns regarding the use of already published movement based indices, and enable us to recommend how new foraging data should be collected.

## Introduction

The study of animal foraging behaviour has been a key topic in ecology and evolutionary biology (Reilly, McBrayer & Miles 2007; Stephens, Brown & Ydenberg 2007). Comparison of foraging behaviour across species and in conjunction with other traits requires condensing the myriad of searching, pursuit, and capture behaviours to simple and comparable quantitative indices (Reilly, McBrayer & Miles 2007). This is probably why simple movement-based indices, and especially Moves Per Minute (MPM) and Proportion Time Moving (PTM), have become very popular, and are still being utilized extensively to characterized foraging behaviour across taxa (e.g. Lizards: Reilly, McBrayer & Miles 2007; Halperin, Carmel & Hawlena 2017; Fish: Radabaugh 1989; Davis, Spencer & Ottmar 2006; Fu *et al.* 2009; Snakes: Hansknecht & Burghardt 2010; Insects: Ferris & Rudolf 2007; Mundahl & Mundahl 2015). For example, the foraging behaviour of approximately 170 lizard species has been characterized using PTM and MPM (see Halperin, Kalyuzhny & Hawlena (2018) for a detailed account).

Huey & Pianka (1981) were the first to introduce PTM and MPM. They recorded the "duration of each move and duration of each stop" of seven Kalahari lizard species, and divided the cumulative duration of all movements or the number of discrete movements by the total observation duration for estimating PTM and MPM, respectively. The authors did not explain how move and stop were defined, and did not specify the minimal durations of the moves and stops they used ($M_{min}$ and $S_{min}$, respectively). This potentially important information has rarely been reported in hundreds of succeeding studies, despite technological advancements that allow recording foraging behaviour more precisely (but see Hawlena *et al.* 2006).

Recently, Halperin, Kalyuzhny & Hawlena (2018) raised a concern that variation in $M_{min}$ and $S_{min}$ may seriously affect MPM and PTM estimates. They suggested that different $M_{min}$ and $S_{min}$ used to characterize exactly the same movement sequence could yield different MPM estimates. This is because as $M_{min}$ and $S_{min}$ decrease, more movement events are included in the calculation, increasing MPM. This problem is expected to affect also the PTM calculation because as $S_{min}$ increases less stop events are included in the calculation, increasing the cumulative time spent moving. Furthermore, inconsistent determination of $M_{min}$ and $S_{min}$ both within and between studies may inflate the inaccuracy and imprecision of the MPM and PTM estimates, possibly affecting case studies and comparative studies alike. An important next step is thus to empirically characterize the degree in which movement based indices are dependent upon $M_{min}$ and $S_{min}$.

To address this requisite we recorded the foraging behaviour of 20 Bosc's fringe-toed lizard (*Acanthodactylus boskianus* Daudin 1802) in the field, using high speed imagining (120 frames per seconds). After digitizing the data, we gradually decreased the resolution by ignoring short stops and either ignoring, elongating or leaving short moves unchanged. We used the manipulated data to conduct a sensitivity analysis that examines to what extent estimates of PTM and movement frequency rely on $M_{min}$ and $S_{min}$. To characterize the movement frequency, we followed the recommendation of Halperin, Kalyuzhny & Hawlena (2018), and used Changes Per Minute (CPM) instead of the classic MPM. This frequency index is very similar to MPM but does not suffer from bias and inflated error (Halperin, Kalyuzhny & Hawlena 2018). CPM is calculated by dividing the number of initiations of stops and moves by the total observation duration (minus one time unit, since no changes can be observed after it). Thus, CPM/2 approximately equals MPM, making these indices fully comparable.

A possible way to reduce variation in the determination of $M_{min}$ and $S_{min}$ is to record the animal behaviour more precisely, and to calculate the indices using the shortest movement and stop durations. Yet, this approach may artificially inflate PTM and CPM estimates. This is because animal locomotion is likely a product of several biological processes; hence not all moves and stops are necessarily relevant for quantifying foraging behaviour. In other words, the relevant minimal move and stop durations may be higher than the absolute $M_{min}$ and $S_{min}$ values. Consequently, our second goal was to search for statistically sound and biologically relevant criteria for determining $M_{min}$ and $S_{min}$.

## Methods

### Study site and species

We observed 7 adults and 13 hatchling Bosc's fringe-toed lizard *(Acanthodactylus boskianus)* in three field sites in the central Negev desert, Israel ($30^0 42'$N $34^0 46'$E). All sites were characterized by flat fluvial beds scattered with small perennial shrubs, mostly *Hammada scoparia* and *Artemisia sieberi*. A previous study estimated that *A. boskianus* adults have MPM of 2.66 and PTM of 0.39, and hatchling MPM of 2.15 and PTM of 0.29 (Hawlena 2009).

### Data collection and digitization

We observed the lizards foraging behaviours in the field during September 2016 and August 2017, when adult and hatchling lizards co-occurred. All observations were conducted by the same observer (TH) between 08:30 and 12:00 AM and between 15:00 and 17:30. Lizards were located by random search. When a lizard was spotted, the observer started recording its behaviour at 120 frames per second, using a Panasonic HC-W850 camcorder, mounted on a tripod. While recording, the observer stood motionless and kept a distance of 3-5 meters from the lizard. If the lizard moved out of sight, the observer slowly moved the camera to a better spot, keeping the stream of recording. All observations were 20-25 minutes long. During data analysis, we ignored observations in which lizards were engaged in social interactions (e.g., chasing conspecifics) or thermoregulation (e.g., staying motionless for long periods in the shade). The first three minutes of every observation were deleted to discard the habituation time of both lizard and observer.

We used the Behavioural Observation Research Interactive Software (BORIS ; Friard & Gamba 2016) to digitize the behavioural data. Videos were replayed on the computer at low speed and the time of beginning and ending of each event was recorded. We defined moving events as any relocation of the lizard's centre of mass. Thus, during a stop event a lizard is not necessarily motionless, and can handle a food item or wave its arms. We also recorded the beginning and ending of periods in which the lizard was hidden, i.e. out of frame or not visible due to a visual obstruction.

**Data analysis**

To study the effect of $M_{min}$ and $S_{min}$ on PTM and CPM estimates, we artificially reduced the resolution of the original data so that short stops and/or moves were not used for the calculation. To simplify the analysis, we used a single parameter for the temporal resolution of the artificial data, denoted by $\tau$. The way short moves and stops are being recorded can differ dramatically within and between studies. This is because the observers may vary in skills and recording methods. Since it is hard to know exactly how an observer will treat short stops and moves, we simulated different scenarios using three different schemes of "data degradation". Based on our personal field experience, we assumed that it is easier to notice and record short moves than short stops. Thus, in all schemes, stops shorter than $\tau$ were "missed" and incorporated into adjacent moves, but short moves were treated differently using three schemes:

a) leaving short moves unchanged. This scheme assumes that the observer can record even short moves but not short stops. This scheme is simplistic, but it allows directly relating the dependence of CPM on $\tau$ to the distribution of stop durations (more on that below).
b) omitting all moves shorter than $\tau$. This scheme assumes the observer deliberately ignored or missed short moves.
c) lengthening moves shorter than $\tau$ into length $\tau$. The elongated moves begin at the same time as the original move but ends later. Since this elongation can generate new stops shorter than $\tau$, any stop that is now shorter than $\tau$ is transformed into movement time. This process of transforming stops into moves and elongating moves is repeated until the movement sequence does not change any more, which happened when there are no moves or stops shorter than $\tau$. This scheme assumes that short movements are easy to detect, but take more time to record.

All three schemes were applied to the 20 individuals in $\tau$ intervals of 0.01 seconds, and PTM and CPM were calculated for each output sequence and plotted against $\tau$. Periods when the individual was hidden were removed from the total observation time, and we did not consider any possible changes (initiation or end of movement) that could occur in that period.

It is interesting to note that a visual inspection of the dependence of CPM on $\tau$ can give us important information on the nature of the distribution of stop durations. As $\tau$ is increased, more and more stops are discarded and moves are merged, reducing CPM by the proportion of stops that are shorter than $\tau$. Hence, the dependence of CPM on $\tau$ is simply:

(1) $CPM(\tau) \approx CPM_0(1-F_{t=\tau})$,

Where $CPM_0$ is the CPM when no stops are discarded and $F_{t=\tau}$ is the cumulative distribution function of the stops durations evaluated at $\tau$. The formula above is not exact mostly because of missing periods in the data. This formula is true only for scheme a.

Animal locomotion is likely a result of several biological processes, suggesting that not all stops are clearly relevant for quantifying foraging behaviours. Consequently, our second goal was to search for statistically sound criteria for determining the appropriate resolution. We used the bout criterion (Sibly, Nott & Fletcher 1990), denoted here by $\theta^*$, to separate stop intervals within bouts of movement, from stop intervals between them. The classical approach for identifying bout criteria is to separate the distribution of stop durations into components, often by fitting a mixed distribution (e.g., using log-survivorship analysis, Slater & Lester 1982; log-likelihood analysis of mixed exponential distributions, Langton, Collett & Sibly 1995, or log-transformation of the intervals, and fitting a mixture of Gaussians, Tolkamp & Kyriazakis 1999). We attempted a slightly different approach. In line with the suggestion of Clauset, Shalizi & Newman (2009), we fitted (using maximum likelihood methods) only the long stops, above a cut-off duration $\theta$, to several candidate distributions – Pareto (power law), Weibull, gamma and lognormal. Stops shorter than $\theta$ were thus discarded. To find the cut-off point, we fitted each distribution over a range of $\theta$s, from 0.01 seconds to 3 seconds in intervals of 0.01 seconds. For each fit we then calculated the Kolmogorov-Smirnov (K-S) goodness of fit statistic. If the long moves (above $\theta^*$) indeed resemble one of these distributions and the short ones do not, we would expect, as suggested by Clauset, Shalizi & Newman (2009), that as $\theta$ increases, the K-S values would rapidly decrease to a minimum around $\theta^*$ and then slowly increase. This is because the very short moves belong to a different distribution, and removing them from the fit by increasing the threshold $\theta$ strongly improves the goodness of fit. Increasing $\theta$ further, beyond $\theta^*$, results in a decrease of sample size, reducing goodness of fit and increasing K-S. Hence, the $\theta$ corresponding to the minimal K-S value can be regarded as an estimator of $\theta^*$. This optimization procedure was performed for all four candidate distributions, and we chose the $\theta^*$ of the distribution with the lowest value of K-S.

Our estimate of $\theta^*$ cannot be reliable if the fit of the distribution is poor. To test distribution fit, we fitted for each individual the four distributions using $\theta^*$, compared their relative goodness of fit using AIC weights and tested whether they are plausible models using a K-S test. Finally, the above analyses assume that stops are independent. To test this assumption, we tested the lag 1 autocorrelation of stop durations. All analyses were performed using Matlab 2016a.

## Results

*Acanthodactylus boskianus* were motionless most of the observation time, at times for very long periods (>100 sec.). Their movements were highly fragmented by many frequent short stops (Fig 1). Considering all individuals, the longest continuous movement recorded was 2.9 sec., and only 0.12% of movements were > 2 sec. The longest stop that was recorded was 532 sec., while 63% of stops were < 0.2 sec., and 81% of the stops were < 0.5 sec. This behaviour led to very high values of CPM - the mean over lizards was 57.9 ± 4.7 (SE). There were very few (<0.2%) short moves and stops with duration on the order of a single frame, hence, the true $S_{min}$ and $M_{min}$ are likely close to 1/120 second. Moreover, no difference in PTM (t test, P=0.388) or CPM (t test, P = 0.115) was found between juvenile and adult lizards.

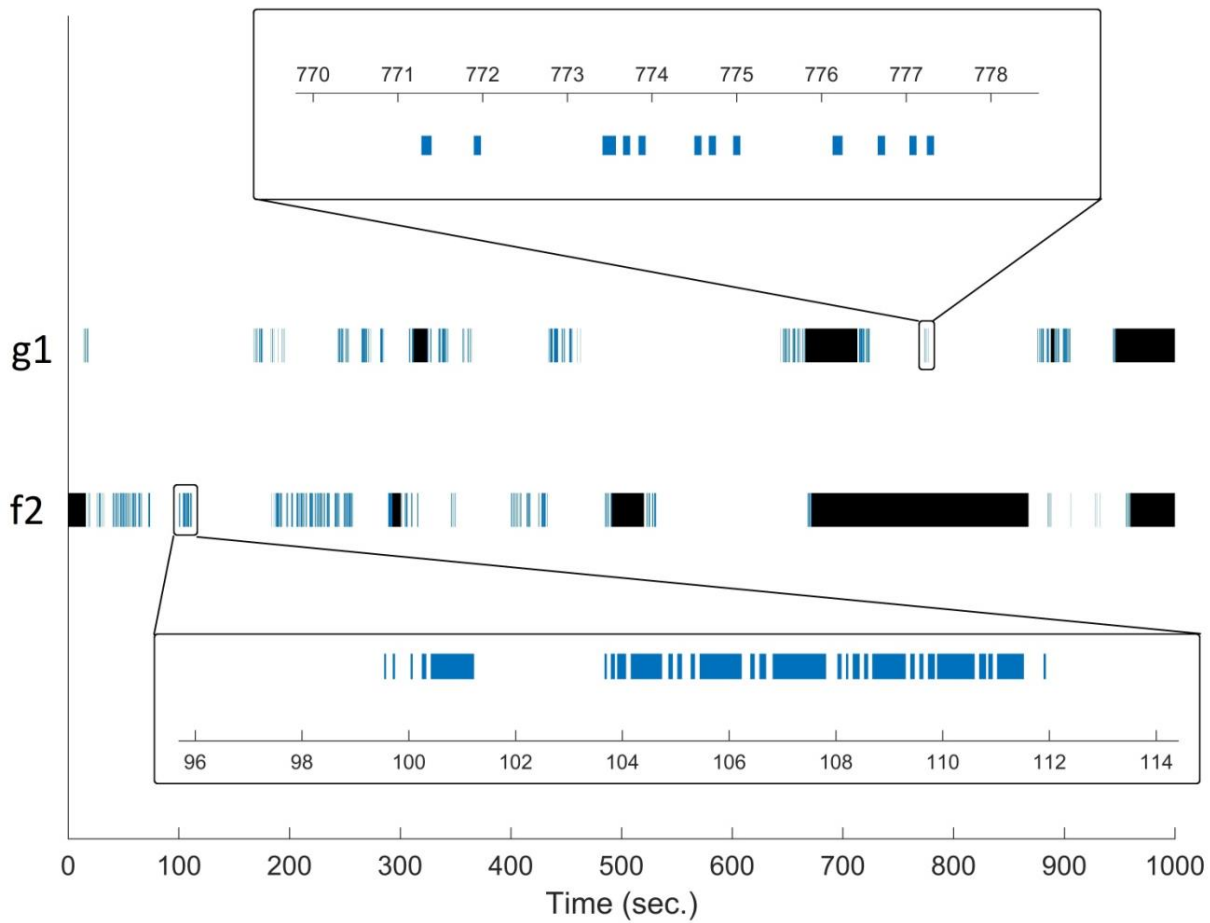

Fig. 1 – Representative movement patterns of two *A. boskianus* lizards (coded "f2" and "g1"). The movement periods (blue), stopping periods (white) and periods when the animal was hidden (black) are presented for 1000 seconds. We also focus on two short time periods, one with few moves and one with more moves. It is evident that there is a very wide distribution of stops; most are very short while some last > 100 sec. On the other hand, there are no long continuous moves.

We found that, regardless of the methodology used to degrade the resolution of the original data, both CPM and PTM had a strong dependence on $\tau$. Fig. 2 presents the simple case when we transformed stops shorter than $\tau$ into moves and left the moves shorter than $\tau$ unchanged (scheme a). In this case, when $\tau$ increased from 0.5 sec. to 3 sec. (which is approximately the relevant range for $\tau$ in field studies), average PTM increased from 0.156 to 0.231 (48% increase), and an increase from 1 sec. to 3 sec. resulted in 29% increase in PTM (Fig. 2a). The effect on CPM was even more dramatic: an increase in $\tau$ from 0.5 sec. and 1 sec. to 3 sec. resulted in a decrease in average CPM from 10.91 to 3.51 (by 68%) and from 6.97 to 3.51 (by 50%), respectively (Fig. 2b). We found similar trends when applying the other two data degradation schemes (Fig. S1-S2).

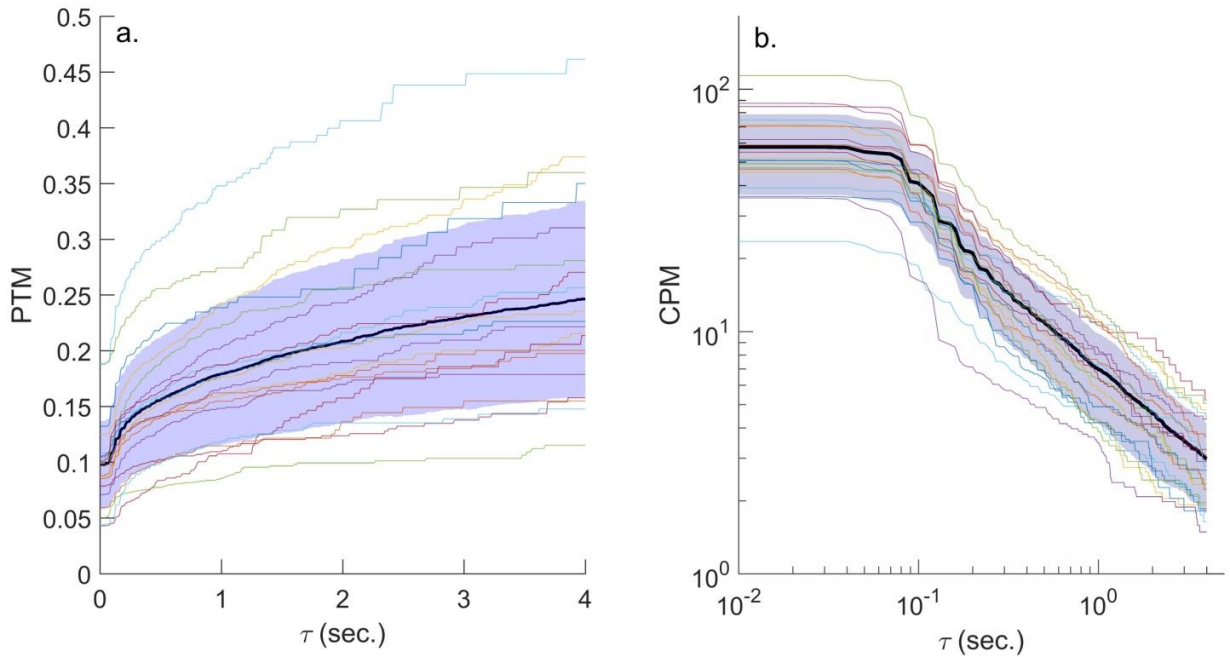

Fig 2 – Dependence of foraging indices on the minimal durations of observed stops, $\tau$, using scheme a). Starting with the original movement sequence of the 20 *A. boskianus* individuals, all stops shorter than $\tau$ were not observed and transformed into movement time, and PTM (a) and CPM (b) were calculated for the new sequences. Each coloured curve presents one individual, while the black line is the average and the blue region is 1 SD around this average. Note the double logarithmic scale of b.

Examining Fig. 2b hinted that the distribution of long stops is wide (possibly a power-law), due to the straight line on the double logarithmic axes, and it seemed that this straight line turns into a different shape for short stops. This was corroborated by our analysis of the distribution of stop durations. The average value of the crossover $\theta^*$ for all lizards was 0.35 ± 0.13 sec. (see Fig. S3 for the search results and table S1 for the values). The search for the crossover $\theta^*$ at the individual level was not very reliable due to the limited number of stops each lizard performed. The lognormal and power-law distributions were indeed plausible distributions for the stop durations: their deviation from the empirical distribution of stops was significant in zero and two individuals (α = .05), respectively. Furthermore, in 18 of the 20 lizards, one (or both) of these distributions was much more plausible than the other candidates, considering AIC weights (Table S1).

The lag 1 autocorrelation of stop durations was not significantly different than zero for all 20 individuals, justifying our treatment of stop events as independent.

## Discussion

We have empirically shown that the two most popular movement based indices for quantifying foraging behaviour, CPM (or MPM) and PTM are very sensitive to the determination of the minimal duration of moves and stops. This result holds even when considering just the range of minimal durations that are relevant for field studies using unassisted vision (~0.5 - 3 sec.). When using the highest resolution, our estimate of CPM was an order of magnitude higher than all prior reported values for lizards. We have also uncovered a crossover point around minimal stop duration of 0.35 sec. that may reveal the relevant resolution for quantifying the foraging behaviour of *A. boskianus*.

Hundreds of field studies have quantified animal foraging behaviour using the two most popular movement-based indices PTM and MPM (Halperin, Kalyuzhny & Hawlena 2018)). These indices are based on the number and duration of moves and stopes. Nevertheless, the temporal resolution of these observations has rarely been reported, despite the possibility that different values of $S_{min}$ and $M_{min}$ may result in very different PTM and MPM estimates. We have shown that for a lizard with intermittent locomotion, PTM and MPM are very sensitive to the minimal duration of moves and stops.

Qualitatively, the existence of such dependence is rather trivial. If short moves and stops are missed then MPM must be smaller, and when short stops are not observed or short moves are recorded as longer moves, PTM must increase. Yet, we were genuinely surprised by the magnitude of this effect. When comparing the highest resolution to a 1 sec. resolution. the estimate of PTM increased by 83% and the estimate of CPM decreased by 88% (considering scheme a).When using the highest resolution, our estimate of CPM (averaged over the individuals and equivalent to MPM≈28.9) was an order of magnitude greater than the highest MPM estimate for lizard reported thus far (i.e., *Psammodromus hispanicus MPM=4.71;* Verwaijen & Van Damme 2007*; see* Halperin, Kalyuzhny & Hawlena (2018) for comparison)*.*

We attributed this unexpected result to the distribution of stop durations. We found that above a threshold of approximately 0.35 sec., the stop durations are well-described by a single, heavy tailed distribution with high kurtosis, resembling a lognormal or a Power law distribution. In such distributions, the vast majority (and the mode) of the stops are very short, but longer stops still occur, and there is no typical time scale above which stops become rare (in the range 0.35 – 4 sec.). Hence, when $\tau$ is increased, many very short stops are incorporated into movement time, considerably increasing PTM. Longer stops are still quite common, and therefore we observe a steady increase in PTM with $\tau$. The strong drop in CPM with $\tau$ occurs for similar reasons – the high frequency of short stops leads, when they are ignored, to a drop in CPM, and because there are also a considerable number of longer stops, this drop is not limited to only a small range of $\tau$.

It is impossible in retrospect to assess what range of minimal stop duration values has been used for calculating MPM and PTM. Based on our field experience, we assumed that a range of minimal stop durations from 0.5 - 3 sec. should encapsulate the realistic variation used in previous field studies. We found a 48% difference in PTM and 68% in CPM estimations when comparing those two $S_{min}$ values. If our results are representative at least of species with intermittent locomotion, then the variation in the determination of $S_{min}$ and $M_{min}$ both within and between studies should inflate inaccuracy and imprecision in the published CPM and PTM estimates. These large biases, which are on the order of the measured values, could seriously impair analyses and conclusions that are based on these data. Moreover, such unaccounted source of variation may seriously question the way these indices have been used and interpreted in dozens of comparative studies during the last 3 decades (Reilly, McBrayer & Miles 2007; Scales & Butler 2016; Halperin, Carmel & Hawlena 2017). For example, several investigators have provided empirically derived boundaries for classifying lizard foraging modes. Cooper & Whiting (1999) found that sit-and-wait foragers exhibited PTM < 0.15 while lizards above this cut-off are active foragers. Our results (using scheme a) showed that when $S_{min}$ increased from 0 sec. to 0.35 sec. and to 2 sec., the PTM value increased from 0.0978 to 0.1475 and to 0.2087, respectively. Hence, changes in $\tau$ could qualitatively alter the foraging mode classification of this species.

The strong dependence of CPM and PTM on $\tau$ essentially raises the question: what value of $\tau$ should be used to quantify foraging behaviour? The distribution of stop durations can be described as a mixture of a lognormal or power law above a crossover $\theta^*$ and by a considerably different distribution below $\theta^*$. A reasonable explanation for these finding is that stop events below and above $\theta^*$ are generated by two different processes. Using the common interpretation of bout criteria (Langton, Collett & Sibly 1995; Tolkamp & Kyriazakis 1999), we assumed that $\theta^*$ distinguishes between longer stops that separate move bouts and short stops that happened within a movement bout, and may not be informative for quantifying foraging behaviour. We suggest that short stops under $\theta^*$ that occurs within a longer movement bout may cause motion dazzle that, together with the longitudinal dorsal stripes of *A. Boskianus*, may impair the predator ability to intercept the moving prey (Halperin, Carmel & Hawlena 2017). Yet even if one interprets the meaning of this threshold value differently, our findings suggest that setting the minimal stop value arbitrarily above this threshold has no statistical justification. Hence, we suggest discarding stops (and possibly moves, as we have shown, the results do not strongly depend on the way movement events are treated) below this value and taking into consideration all moves and stops above it.

To encourage better use of movement based indices for quantifying foraging behaviour we next discuss the use of already published data, and recommend a new protocol for collecting and calculating PTM and CPM in the field. Published data should be treated with great caution because the published values heavily rely on the arbitrary ways in which $S_{min}$ and $M_{min}$ have been determined. This problem is likely more severe for species with intermittent mobility that exhibits short-frequent moves and stops. Species with low mobility are less likely to perform short-frequent moves and stops because this behaviour may interfere with their sit-and wait strategy. Thus, published PTM (and even CPM) estimates for such species are expected to be more reliable. Highly mobile species may use multiple short stops, but their effect on PTM (relative to mean PTM) is expected to be less pronounced because their PTM is high anyway and transforming some short stops into additional movement time most likely would not create a strong effect. Generally, published data can be trusted if there is a compelling reason to believe that the animal does not perform frequent short stops. Short field observations may assist determining this point.

When new data is collected we recommend recording the foraging behaviour with a high-speed camera, encoding the movies after clearly defining movement, and then analysing the distribution of stops in order to find the bout criterion above which all stops are clearly relevant for quantifying foraging. Finally, we suggest calculating PTM and CPM based on all moves and stops that are greater than the bout criterion, and reporting the details of this procedure to enhance comparability and reproducibility.

In summary, movement-based indices such as PTM and MPM are common methodologies to quantify foraging behaviour that are highly advantageous for comparative evolutionary-ecological studies. Our work suggest that the ways foraging data have been collected, analysed and interpreted in hundreds of studies is potentially flawed. By precisely recording the foraging behaviour of *A.boskianus* lizards and artificially reducing the resolution of these observations, we showed that in this lizard the determination of the minimal stops and moves substantially affect the PTM and CPM estimates. If these findings hold for other species than the use of published PTM and MPM estimates in comparative studies must be treated with extra caution. Nevertheless, we believe that with the lack of better alternatives, the use of simple and informative statistics such as PTM and CPM for comparative studies will persist in the foreseen future. Thus, we hope that our proposed procedure for identifying the minimal moves and stops, together with the recent recommendations of how to use (and not to use) movement based foraging indices (Halperin, Kalyuzhny & Hawlena 2018), will add rigor to the way foraging data is being collected, analysed and used.

## Acknowledgements

We acknowledge Topaz Halperin for her invaluable comments suggestions, the members of the risk management ecology lab for stimulating discussions, and the support of a European Research Council grant (ERC-2013-StG-337023 (ECOSTRESS)) to D.H. M.K. is supported by the Adams Fellowship Program of the Israel Academy of Sciences and Humanities.

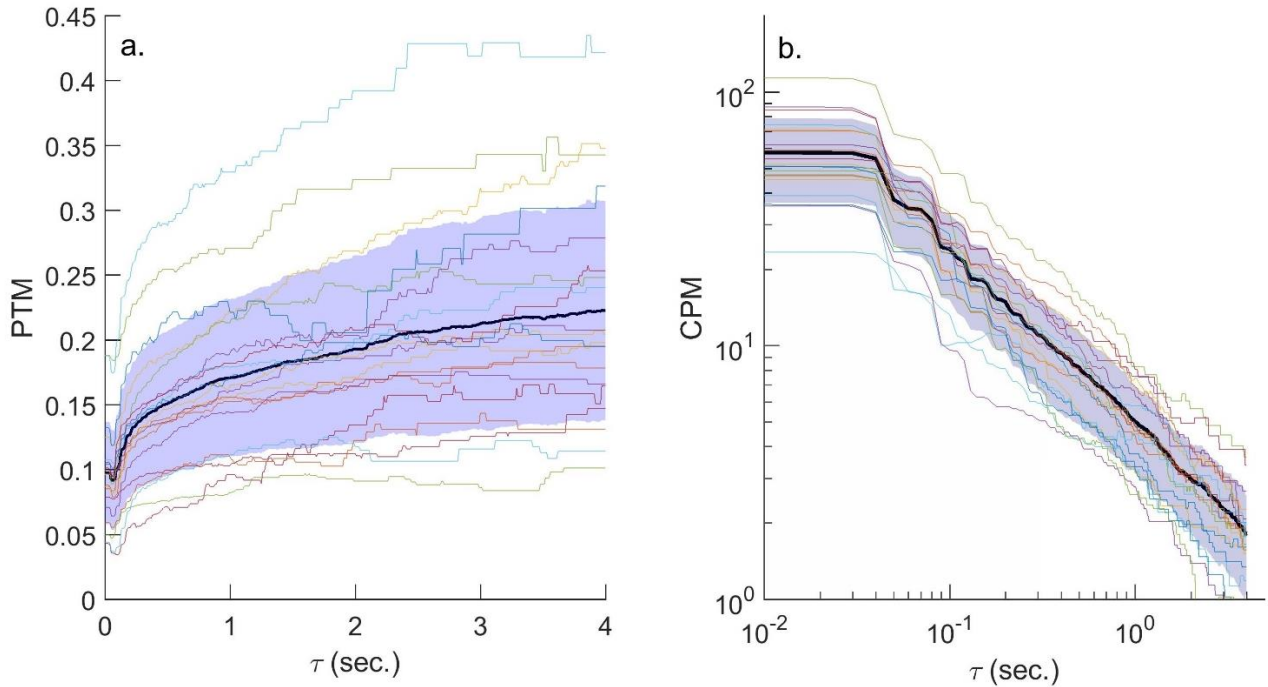

Fig S1 – Dependence of foraging indices on the minimal durations of observed stops and moves, $\tau$, using scheme b). Starting with the original movement sequences of the 20 *A. boskianus* individuals, all stops shorter than $\tau$ were not observed and transformed into movement time, and following that all moves shorter than $\tau$ were not observed and transformed into stopping time. PTM (a) and CPM (b) were calculated for the new sequences. Each colored curve presents one individual, while the black line is the average and the blue region is 1 SD around this average. Note the double logarithmic scale of b.

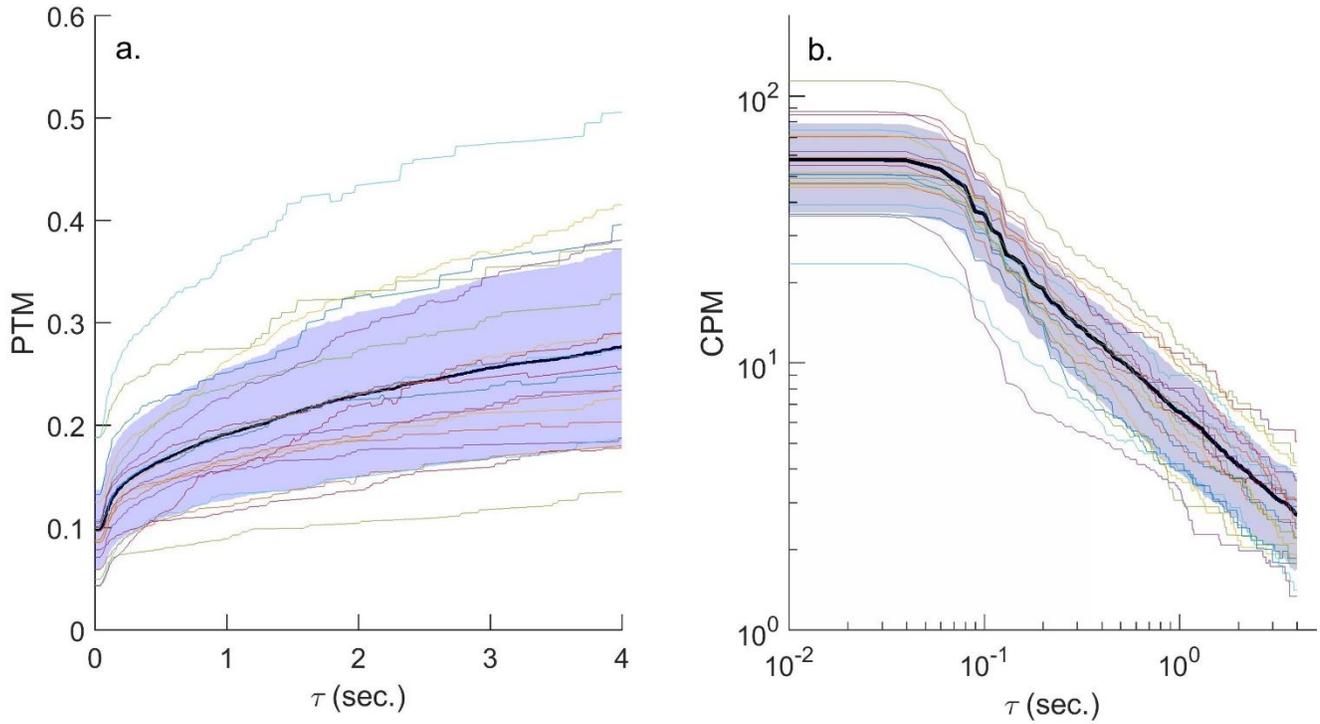

Fig S2 – Dependence of foraging indices on the minimal durations of observed stops and moves, $\tau$, using scheme c). Starting with the original movement sequence of the 20 *A. boskianus* individuals, all stops shorter than $\tau$ were not observed and transformed into movement time, and following that all moves shorter than $\tau$ were elongated forward in time and transformed into moves of time $\tau$. These two steps were repeated until the sequence did not change further. PTM (a) and CPM (b) were then calculated for the new sequences. Each colored curve presents one individual, while the black line is the average and the blue region is 1 SD around this average. Note the double logarithmic scale of b.

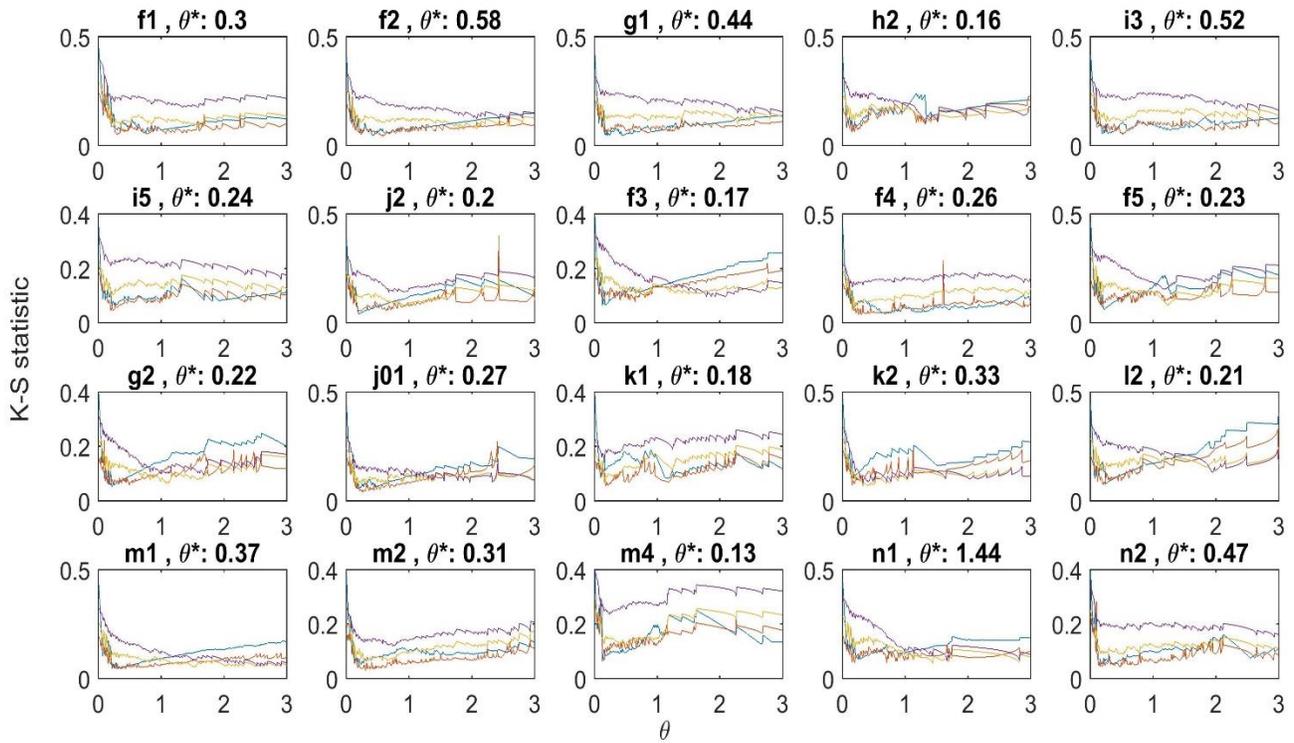

Fig. S3 - Estimation of the crossover ($\theta^*$) between short and long stops in the distribution of stop durations for 20 *A. boskianus* indivisuals. For values of the crossover parameter $\theta$ between 0 sec. and 3 sec., all stops shorter than $\theta$ were discarded, and the other stops were fitted using maximum likelihood to the pareto (blue), lognormal (red), weibull (orange) and gamma (purple) distributions. The K-S statistic of these distributions as a function of $\theta$ is presented. In the title we show the code of the individual and the value of $\theta$ that minimizes K-S, $\theta^*$.

| name | Pareto (power law) | | | | Lognormal | | | | | Weibull | | | | | Gamma | | | | | θ* |
|---|---|---|---|---|---|---|---|---|---|---|---|---|---|---|---|---|---|---|---|---|
| | α | K-S | Pval. | AIC$_w$ | μ | σ | K-S | Pval. | AIC$_w$ | a (scale) | b (shape) | K-S | Pval. | AIC$_w$ | a (shape) | b (scale) | K-S | Pval. | AIC$_w$ | |
| 'f1' | 0.583 | 0.073 | 0.801 | 0.503 | -0.160 | 2.180 | 0.051 | 0.986 | 0.497 | 2.596 | 0.438 | 0.107 | 0.336 | 0.000 | 0.275 | 39.624 | 0.217 | 0.001 | 0.000 | 0.300 |
| 'f2' | 0.637 | 0.075 | 0.812 | 0.131 | 0.144 | 2.390 | 0.045 | 0.998 | 0.850 | 3.797 | 0.439 | 0.106 | 0.391 | 0.019 | 0.298 | 39.531 | 0.188 | 0.013 | 0.000 | 0.580 |
| 'g1' | 0.661 | 0.048 | 0.980 | 0.921 | -0.025 | 2.067 | 0.086 | 0.460 | 0.079 | 2.854 | 0.456 | 0.134 | 0.060 | 0.000 | 0.305 | 30.530 | 0.236 | 0.000 | 0.000 | 0.440 |
| 'h2' | 0.748 | 0.102 | 0.465 | 0.375 | -1.449 | 2.230 | 0.065 | 0.922 | 0.625 | 0.735 | 0.432 | 0.112 | 0.335 | 0.000 | 0.277 | 10.587 | 0.224 | 0.002 | 0.000 | 0.160 |
| 'i3' | 0.735 | 0.112 | 0.204 | 0.069 | -0.295 | 2.343 | 0.053 | 0.956 | 0.930 | 2.409 | 0.435 | 0.110 | 0.213 | 0.001 | 0.287 | 29.178 | 0.211 | 0.001 | 0.000 | 0.520 |
| 'i5' | 0.545 | 0.081 | 0.654 | 0.260 | -0.336 | 2.471 | 0.046 | 0.993 | 0.739 | 2.505 | 0.401 | 0.102 | 0.352 | 0.000 | 0.253 | 46.566 | 0.218 | 0.001 | 0.000 | 0.240 |
| 'j2' | 0.627 | 0.042 | 0.981 | 0.697 | -0.727 | 2.134 | 0.055 | 0.834 | 0.303 | 1.441 | 0.446 | 0.118 | 0.060 | 0.000 | 0.285 | 19.363 | 0.231 | 0.000 | 0.000 | 0.200 |
| 'f3' | 0.649 | 0.066 | 0.564 | 0.999 | -0.987 | 2.251 | 0.110 | 0.063 | 0.001 | 1.183 | 0.418 | 0.170 | 0.001 | 0.000 | 0.262 | 21.094 | 0.274 | 0.000 | 0.000 | 0.170 |
| 'f4' | 0.661 | 0.103 | 0.022 | 0.002 | -0.420 | 1.729 | 0.039 | 0.898 | 0.998 | 1.592 | 0.538 | 0.099 | 0.030 | 0.000 | 0.380 | 9.936 | 0.193 | 0.000 | 0.000 | 0.260 |
| 'f5' | 0.775 | 0.062 | 0.699 | 0.986 | -1.066 | 2.111 | 0.096 | 0.184 | 0.014 | 1.054 | 0.410 | 0.166 | 0.002 | 0.000 | 0.238 | 29.217 | 0.306 | 0.000 | 0.000 | 0.230 |
| 'g2' | 0.707 | 0.053 | 0.614 | 0.919 | -0.854 | 1.993 | 0.079 | 0.158 | 0.081 | 1.213 | 0.468 | 0.139 | 0.001 | 0.000 | 0.320 | 11.175 | 0.232 | 0.000 | 0.000 | 0.220 |
| 'j01' | 0.683 | 0.073 | 0.813 | 0.178 | -0.534 | 1.855 | 0.042 | 0.999 | 0.813 | 1.486 | 0.563 | 0.095 | 0.490 | 0.008 | 0.432 | 6.116 | 0.155 | 0.051 | 0.000 | 0.270 |
| 'k1' | 0.536 | 0.116 | 0.133 | 0.011 | -0.333 | 1.882 | 0.055 | 0.917 | 0.989 | 1.859 | 0.514 | 0.092 | 0.351 | 0.000 | 0.358 | 13.065 | 0.187 | 0.002 | 0.000 | 0.180 |
| 'k2' | 0.583 | 0.147 | 0.377 | 0.003 | -0.318 | 2.634 | 0.118 | 0.636 | 0.017 | 2.400 | 0.493 | 0.068 | 0.991 | 0.521 | 0.362 | 12.784 | 0.081 | 0.953 | 0.459 | 0.330 |
| 'l2' | 0.718 | 0.167 | 0.046 | 0.035 | -1.269 | 2.580 | 0.076 | 0.816 | 0.965 | 1.067 | 0.369 | 0.137 | 0.150 | 0.000 | 0.224 | 31.883 | 0.253 | 0.000 | 0.000 | 0.210 |
| 'm1' | 0.616 | 0.052 | 0.764 | 0.041 | -0.172 | 2.316 | 0.043 | 0.921 | 0.957 | 2.635 | 0.464 | 0.101 | 0.070 | 0.002 | 0.325 | 21.088 | 0.169 | 0.000 | 0.000 | 0.370 |
| 'm2' | 0.630 | 0.069 | 0.484 | 0.027 | -0.235 | 1.939 | 0.030 | 0.999 | 0.973 | 2.076 | 0.533 | 0.067 | 0.517 | 0.000 | 0.390 | 11.062 | 0.142 | 0.005 | 0.000 | 0.310 |
| 'm4' | 0.538 | 0.065 | 0.795 | 0.857 | -0.771 | 2.226 | 0.085 | 0.465 | 0.143 | 1.492 | 0.396 | 0.148 | 0.027 | 0.000 | 0.224 | 51.666 | 0.273 | 0.000 | 0.000 | 0.130 |
| 'n1' | 0.490 | 0.145 | 0.764 | 0.020 | 1.602 | 2.773 | 0.123 | 0.888 | 0.033 | 16.776 | 0.492 | 0.069 | 1.000 | 0.444 | 0.360 | 88.381 | 0.102 | 0.971 | 0.504 | 1.440 |
| 'n2' | 0.677 | 0.067 | 0.758 | 0.355 | -0.073 | 2.119 | 0.043 | 0.991 | 0.645 | 2.695 | 0.474 | 0.091 | 0.373 | 0.000 | 0.322 | 24.100 | 0.188 | 0.002 | 0.000 | 0.470 |

Table S1 – Results of the fits of stop duration distributions (in seconds) of 20 *A. boskianus* lizards. For each individual, we present the crossover that was found between long and short stops, $\theta^*$, as well as the properties of the fits of stops longer than $\theta^*$ to four candidate distributions – Pareto, Lognormal, Weibull and Gamma. For each distribution we present the fitted parameters, the Kolmogorov-Smirnoff statistic (K-S) and its corresponding P-value, as well as the Akaike weight when the four different possible distributions are considered. The parameters of the Lognormal are the mean and SD of the corresponding normal distribution.